\documentclass[american,aps,pra, reprint]{revtex4-1}
\pdfoutput=1
\usepackage[T1]{fontenc}
\usepackage[latin9]{inputenc}
\usepackage{babel}
\usepackage{amsthm}
\usepackage{amsmath}
\usepackage{bm}
\usepackage{amssymb}
\usepackage{graphics,graphicx}
\usepackage{tikz}
\usepackage{float}
\usepackage[unicode=true,pdfusetitle, bookmarks=true,bookmarksnumbered=false,bookmarksopen=false,
breaklinks=false,pdfborder={0 0 0},backref=false,colorlinks=false]
{hyperref}
\hypersetup{colorlinks,linkcolor=myurlcolor,citecolor=myurlcolor,urlcolor=myurlcolor}

\makeatletter
\@ifundefined{textcolor}{}
{%
	\definecolor{BLACK}{gray}{0}
	\definecolor{WHITE}{gray}{1}
	\definecolor{RED}{rgb}{1,0,0}
	\definecolor{GREEN}{rgb}{0,1,0}
	\definecolor{BLUE}{rgb}{0,0,1}
	\definecolor{CYAN}{cmyk}{1,0,0,0}
	\definecolor{MAGENTA}{cmyk}{0,1,0,0}
	\definecolor{YELLOW}{cmyk}{0,0,1,0}
}
\theoremstyle{plain}

\ifx\proof\undefined
\newenvironment{proof}[1][\protect\proofname]{\par
	\normalfont\topsep6\p@\@plus6\p@\relax
	\trivlist
	\itemindent\parindent
	\item[\hskip\labelsep
	\scshape
	#1]\ignorespaces
}{%
	\endtrivlist\@endpefalse
}
\providecommand{\proofname}{Proof}
\fi

\usepackage{times}
\usepackage{txfonts}
\usepackage{braket}
\usepackage{colortbl}
\definecolor{myurlcolor}{rgb}{0,0,0.7}

\makeatother

\providecommand{\theoremname}{Theorem}

\usepackage{times}

\usepackage[up]{subfigure}

\begin{document}
	\title{ Complementarity Relation for Coherence and Disturbance}
	
	\begin{abstract}
		Quantum measurements necessarily disturb the state of physical system. Once we perform a complete measurement, the system undergoes decoherence and loses its coherence. If there is no disturbance, the state retains all of its coherence. It is therefore natural to ask if there is trade-off between disturbance caused to a state and its coherence. We present a coherence disturbance complementarity relation using the relative entropy of coherence. For bipartite states we prove a complementarity relation between the quantum coherence, entanglement and disturbance. Similar relation also holds for quantum coherence, quantum discord and disturbance for a bipartite state. We illustrate the trade-off between the coherence and the disturbance for single qubit state for various quantum channels.  
		
	\end{abstract}
	
	\author{Gautam Sharma}
	
	\author{Arun Kumar Pati}
	
	\affiliation{Quantum Information and Computation Group,\\ \mbox{Harish-Chandra Research Institute, HBNI, Chhatnag Road, Jhunsi, Allahabad 211 019, India}}


	\maketitle

	\section{Introduction}
	Measurement is an integral part of quantum theory, for it gives us information about the physical system. Unlike classical systems, in quantum mechanics the measurement process necessarily disturbs the state of the system unless the state is in one of the eigenstate of the observable being measured. Intuitively, we know that if we want to extract information, the state of the system is necessarily disturbed, however for most information processing tasks one would like to keep the disturbance to be minimum. There are several papers aimed at proving this statement quantitatively by deriving information vs disturbance trade-off relations, using different definitions of information and disturbance in varied scenarios \cite{PhysRevA.53.2038, PROP:PROP200310045, maccone2006information, PhysRevLett.68.557, PhysRevLett.86.1366, PhysRevA.64.052307,  PhysRevLett.96.020408, PhysRevLett.96.020409, PhysRevLett.96.220502, PhysRevA.74.052320, PhysRevLett.100.210504, PhysRevLett.109.150402, PhysRevA.93.022104}. It is well known that a state of a single quantum system cannot be determined if we demand no disturbance and possess no prior information about the quantum state \cite{Wootters, PhysRevLett.76.2832}. However, the disturbance of the quantum system can be made arbitrarily small by using weak measurements. The weak measurement also has potential applications in quantum information processing \cite{AHARONOV199338,PhysRevA.47.4616,Aharonov1996,PhysRevA.81.040103, nphys2178}.
	
	The measurement process not only disturbs the state, but also leads to loss of coherence. This  results in the decoherence of a system which is seen as diminishing of the off-diagonal elements of the density matrix that indicates the loss of superposition. Both direct and indirect (here one uses the ancilla) measurement processes  \cite{PhysRevA.93.012115, RevModPhys.75.715, Petruccione2002} can cause decoherence of system. Essentially, both direct and indirect measurements result in transfer of information from the system to the apparatus and environment. 
	
	Coherence is a property of the physical system in the quantum world that can be used to drive various non-classical phenomena. Hence, coherence can be viewed as a resource, which enables us to perform useful quantum information processing tasks. Much before the resource theory of coherence was developed \cite{PhysRevLett.113.140401,cohresource, PhysRevLett.117.020402, PhysRevA.91.052115}, coherence was viewed as a resource similar to entanglement. In fact, similar to the entanglement swapping, the coherence swapping has been proposed that can create coherent superposition from two incoherent states \cite{Pati2001}. After the development of the resources theory of coherence, this was shown to be complementary to the path distinguishability in an interferometer \cite{PhysRevA.92.012118}.  Similarly, a complementarity relation between quantum coherence and entanglement was proved in Ref.\cite{coherenceentanglement}. Also, coherence in two incompatible basis were shown to be complementary to each other by proving that they satisfy an uncertainty relation in Ref.\cite{math4030047}. Complementarity of coherence with mixedness and asymmetry was also investigated in Refs.\cite{PhysRevA.92.042101,antiunitarity,coherencecomple}.
	
	This motivates us to explore the idea that the initial coherence should respect a trade-off relation with the disturbance caused to the system state, whenever some information is extracted from the system or a measurement is performed. In this work we present a complementarity relation between the initial coherence of the system state and the disturbance caused by a CPTP map. The complementarity relation is tight and the equality is satisfied for amplitude damping and depolarising channels in the case of single qubit state, while the bound is lowered as the measurement strength is reduced. To prove the complementarity relation, we have used a measure of disturbance given in Ref.\cite{0295-5075-77-4-40002} and the relative entropy quantum coherence given in Ref.\cite{PhysRevLett.113.140401}. For a bipartite state we prove a complementarity relation for quantum coherence, entanglement and the disturbance induced by quantum operation. In addition, we also prove a similar relation for quantum coherence, quantum discord and disturbance.
	
	The rest of the paper is organized as follows. In section \ref{sec:2}, we briefly review the basics of quantum measurement and quantification of coherence and disturbance. Next, in section \ref{sec:3}, we derive the complementarity relation between coherence and disturbance which is our main result. For a bipartite state we prove a complementarity relation for the relative entropy of coherence, the relative entropy of entanglement and the disturbance induced by quantum operation. We also prove that a similar relation holds for quantum coherence, quantum discord and disturbance. In section \ref{sec:4}, we give a few examples to illustrate the complementarity between quantum coherence and disturbance for various quantum channels. In section \ref{sec:5}, we present the conclusion and implications of our results.

\section{Basic definitions and preliminaries}\label{sec:2}

\subsection{Quantum Measurement}
Quantum measurement is a distinct type of evolution compared to the Schr\"{o}dinger evolution of a quantum system. The measurement process is non-unitary in nature and gives us classical information about the system. In quantum mechanics one can measure the observables represented by Hermitian operators. After the measurement process the state collapses to one of the eigenstates of the observable with certain probability. This kind of measurement is described using the projection operators $\Pi_j$ with $\sum_j \Pi_j=1$. However, the most general measurement can be described using a set of measurement operators $M_j$ such that $\sum_j M^{\dagger}_j M_j= 1$.  
For an initial state $\rho$, the probability of obtaining an outcome $j$ is $p_j= \mathrm{Tr}(\Lambda_j \rho)$, where $\Lambda_j$ is the apparatus POVM (Positive Operator-Valued Measure). Whenever there is loss in information of measurement outcome, the final state of the system is given by a noisy map $\mathcal{N}(\rho)=\Sigma_j M_j\rho M^{\dagger}_j$. In order that the final state is a proper density matrix operator, the map $\mathcal{N}(\rho)$ should be a CPTP (completely positive trace-preserving) map.

\subsection{Quantifying Coherence and Disturbance}
In this section we briefly review the definitions of coherence  of a quantum system and the disturbance caused to the quantum system due to a quantum operation.

\subsubsection*{Quantum Coherence}
Quantum coherence arises from the superposition principle, thus marks the departure from classical physics.  It is a basis dependent quantity hence it is necessary to fix the reference basis in which we define a quantitative measure of coherence. An axiomatic approach to quantify quantum coherence was developed by Baumgratz et al. in Ref.\cite{PhysRevLett.113.140401} by characterizing incoherent states $\mathcal{I}$ and incoherent operations $\Lambda$. For a given reference basis ${\ket{i}} , {(i=0,1,...d-1)}$, all incoherent states are of the form $\rho=\sum_i p_i \ket{i}\bra{i}$ such that $\sum_i p_i=1$. All incoherent operators are defined as CPTP maps, which map the incoherent states onto itself.   A genuine measure of quantum coherence should fulfill the following requirements: $(i)$	Non-negativity: $C(\rho)\geq 0$ in general. The equality is satisfied iff $\rho$ is an incoherent state. $(ii)$ Monotonicity: $C(\rho)$ does not increase under the action of incoherent operations, i.e., $C(\Lambda(\rho)) \leq C(\rho)$, where $\Lambda$ is an incoherent operation. $(iii)$ Strong monotonicity: $C(\rho)$ does not increase on average under selective incoherent operations, i.e., $\sum_i q_i C(\sigma_i)\leq C(\rho)$, where $q_i = \mathrm{Tr}[K_i \rho {K^\dagger}_i]$ are the probabilities, $\sigma_i = {K_i \rho {K^\dagger}_i}/q_i$ are post measurements states and $K_i$ are the incoherent Kraus operators. $(iv)$ Convexity: $C(\rho)$ is a convex function of the state, i.e., $\sum_i p_i C(\rho_i) \geq C(\sum_i p_i \rho_i)$. It can be noted that conditions $(iii)$ and $(iv)$ put together imply the condition $(ii)$.

The measures that fulfill the above requirements are the $l_1$ norm of coherence and the relative entropy of coherence. In the present work we have used the relative entropy of coherence given by 
\begin{equation}\label{eq:1}
C_r(\rho) = S(\rho^{D})-S(\rho), 
\end{equation}
where $S(\rho)=-\mathrm{Tr}(\rho\log_2(\rho))$ is the von Neumann Entropy of the density matrix $\rho$ and $\rho^D$ denotes the state obtained by deleting the off-diagonal elements of $\rho$.  For a given $d$ dimensional state, $0\leq C_r(\rho) \leq \log_2(d)$. Hence using the above definition we can define maximally coherent state with $C_r(\rho)=\log_2(d)$, which is the case when $\ket{\psi_d}=\frac{1}{\sqrt{d}}{\sum}_{i=0}^{d-1} \ket{i}$.  Another definition of coherence based on matrix norm is the $l_1$ norm of coherence, which is given by  $C_{l_1}(\rho) = \sum_{i\neq j} |\rho _{ij}| $, where $\rho_{ij}=\bra{i}\rho\ket{j}$. Also a geometric measure of coherence was defined in Ref.\cite{PhysRevLett.115.020403}, as $C_g(\rho) = 1-\max_{\sigma \epsilon I}F(\rho,\sigma)$, where $I$ is the set of all incoherent states and the fidelity $F(\rho,\sigma)=||\sqrt{\rho}\sqrt{\sigma}||^2_1$.

\subsubsection*{Disturbance}
In quantum scenario disturbance caused by a measurement process can be defined with respect to both the observable and the state. The disturbance for an observable due to measurement of another observable was defined in Ref.\cite{PhysRevA.67.042105,Ozawa2004350} and for state in Ref.\cite{PhysRevLett.111.160405,PhysRevA.84.042121,PhysRevA.89.022106} to prove the error-disturbance relations. However, here we consider the disturbance caused to a state by the measurement process and do not aim to formulate error-disturbance relations. We say that a system is disturbed when the initial and final state do not coincide. Disturbance is an irreversible change in the state of the system, caused by CPTP evolution. It is thus required that the quantity $D$ that measures disturbance should satisfy the following conditions: $(i)$ $D$ should be a function of the initial state $\rho$ and the CPTP map $\cal{E}$ only, i.e., $D=D(\rho,\mathcal{E})$, $(ii)$ $D(\rho,\mathcal{E})$ should be null iff the CPTP map is invertible on the initial state $\rho$, because in this case the change in state can be reversed hence the system is not disturbed, $(iii)$  $D(\rho,\mathcal{E})$ should be monotonically non-decreasing under successive application of CPTP maps and $(iv)$  $D(\rho,\mathcal{E})$ should be continuous for maps and initial states which do not differ too much.

Several definitions of disturbance have been proposed using the fidelity and the Bures distance between the initial and final state \cite{PhysRevA.53.2038,PROP:PROP535,maccone2006information}, but they fail to satisfy the irreversibilty condition. Moreover, the fidelity based definition is non-zero for unitary transformations which are reversible. Also these definitions can be null for non-invertible maps and they are not monotonically non-decreasing under successive application of CPTP maps, therefore they fail to satisfy conditions $(ii)$ and $(iii)$. It was shown by Maccone that all the above conditions are met by the following definition of disturbance \cite{0295-5075-77-4-40002}:
\begin{align}\label{eq:2}
D(\rho,\mathcal{E}) &\equiv S(\rho) -I_c(\rho) \nonumber \\ &= S(\rho)-S(\mathcal{E}(\rho)) + S((\mathcal{E}\otimes I) (\ket{\Psi}_{SR}\bra{\Psi}))
\end{align}

\noindent where $I_c=S(\mathcal{E}(\rho)) - S((\mathcal{E}\otimes I) (\ket{\Psi}_{SR}\bra{\Psi}))$ is the coherent information  \cite{PhysRevA.54.2629,PhysRevA.56.131} of the system passing through a noisy channel, and $\ket{\Psi}_{SR}\bra{\Psi}$ is a purification of $\rho$, such that $\rho=\rho_S=\mathrm{Tr}_R(\ket{\Psi}_{SR}\bra{\Psi})$. The map $\mathcal{E} \otimes I$ acts on $\ket{\Psi}_{SR}$ with $\mathcal{E}$ acting on the system space and $I$ acts on the ancilla (purification) space. The quantity $I_c$ is non-increasing under successive application of CPTP maps, which makes disturbance monotonically non-decreasing under CPTP maps. It is clear from the definition of $D(\rho,\mathcal{E})$ that for a $d$-dimensional density matrix $\rho$ satisfies, $0 \leq D \leq 2\log_2(d)$.  

\section{Complementarity relations: Coherence, Entanglement, Quantum Correlations and Disturbance}\label{sec:3}

In this section we shall investigate how the initial coherence of the density matrix should respect a trade-off relation with the disturbance caused by a quantum operation. Similarly, for a bipartite state we will explore how the quantum features like coherence, entanglement and quantum discord should respect a trade-off relation with the disturbance caused by a CPTP map. 
\subsection{Coherence-Disturbance complementarity relation}
We prove that there exists, a complementarity relation between the amount of coherence and the disturbance caused to a system by a CPTP map.
Consider a $d$ dimensional system density matrix $\rho$, initially the system and ancilla $\rho_R$ share a pure bipartite state $\ket{\Psi}_{SR}$. The system undergoes a quantum operation $\mathcal{E}$ while there is no quantum processing on the ancilla.
The complementarity relation is given by 
\begin{align}\label{eq:3}
2C_r(\rho) + D(\rho,\mathcal{E}) \leq 2\log(d).
\end{align}
The proof is as follows :
\begin{align*}
	&2C_{rel. ent}(\rho) + D(\rho,\mathcal{E}) \nonumber \\
	&= 2S(\rho^{D})-S(\rho)-S(\mathcal{E}(\rho)) + S(\mathcal{E}\otimes I (\ket{\Psi}_{SR}\bra{\Psi})) \nonumber \\
	&\leq   2S(\rho^{D})-S(\rho)-S(\mathcal{E}(\rho)) + S(\mathcal{E}(\rho)) + S(\acute{\rho_R}) \nonumber  \\
	&=  2S(\rho^{D})-S(\rho) + S(\rho_R) \nonumber \\
	&=  2S(\rho^{D}) \nonumber \\
	&\leq 2\log(d)   \nonumber
\end{align*} 
\noindent where $\acute{\rho_R}$ is the final state of ancilla and the log has base 2. The first inequality is obtained by using the subadditivity of quantum entropy. The next  inequality is obtained using the fact that there is no change in entropy of ancilla and the next equality follows using the fact that initial bipartite state is a pure state thus $S(\rho)=S(\rho_S)= S(\rho_R)$. Final inequality comes from the maximum value of entropy of a state.
From Eq.(\ref{eq:3}) we infer that for a given initial coherence of the state the maximum allowed disturbance lies on the straight line $ D_{max}(\rho,\mathcal{E}) = 2\log(d)-2C_{rel. ent}(\rho) $.

\subsection{Coherence-Disturbance complementarity for the measurement channel}
 While the complementarity relation holds true for all quantum channels, the bound is tighter in the case of measurement channels. The quantum operation for the measurement channel is given by
\begin{align*}
\rho\longrightarrow \mathcal{E}(\rho)=\sum_k\Pi_k\rho\Pi_k
                   =\rho^D=\sum_{k}\rho_{kk}\ket{k}\bra{k},
\end{align*} 

\noindent
where $\Pi_k$ are the projection operators. Now if we consider an environment state $\ket{0}_E$ so that $\ket{\Psi}_{SR}\otimes\ket{0}_E$ is also a pure state, then the evolution
$(\mathcal{E}\otimes I) (\ket{\Psi}_{SR}\bra{\Psi})$ is equivalent to unitary evolution of the tripartite state ($U$ acts on $\mathcal{H}_{S}\otimes\mathcal{H}_E$).
\begin{align*}
U\otimes\mathcal{I}(\ket{\Psi}_{SR}\otimes\ket{0}_{E})\longrightarrow \ket{\Psi^\prime}_{SRE}.
\end{align*}

Since $\ket{\Psi^\prime}_{SRE}$ is also a pure state, we have $S(\rho_{SR}^\prime)=S(\rho_E^\prime)$,
where $\rho_{SR}^\prime=(\mathcal{E}\otimes I) (\ket{\Psi}_{SR}\bra{\Psi})=\mathrm{Tr}[U(\ket{\Psi}_{SR}\bra{\Psi}\otimes \ket{0}_E\bra{0})U^{\dagger}]$. Then, using subadditivity of entropy one can obtain the following complementarity relation
\begin{align}\label{eq:4}
C(\rho)+D(\rho,\mathcal{E})\leq\log d_E,
\end{align}
where $d_E=$dim$(\mathcal{H}_E)$, is the dimension of the Hilbert  space of the environment.

\subsection{Complementarity of coherence, entanglement and disturbance}
In the previous section, we proved the complementarity of coherence and disturbance for a single system. However, when we deal with a composite system
it can have coherence, entanglement and quantum correlation beyond entanglement such as discord.
Then, a natural question to ask here is 
if there exists any complementarity relation between the coherence, entanglement and disturbance caused by CPTP maps. In the same spirit one may ask if there is a complementarity relation for the coherence, quantum correlations and disturbance.  Already, we know that
for pure bipartite states there is a complementarity between the relative entropy of coherence and the bipartite entanglement, i.e., 
$C(\rho_A) + E(\ket{\Psi}_{AB}) \le \log d$, where $d$ is the dimension of the subsystem Hilbert space of $A$  \cite{coherenceentanglement}. Below, we prove that there is indeed a complementarity relation for the coherence, relative entropy of entanglement and disturbance caused by measurement or a CPTP map on the bipartite state.

 Suppose, we have the bipartite state $\rho_{AB}$  with purification $\ket{\Psi}_{ABR}$, such that $\rho_{AB}=\mathrm{Tr}_R(\ket{\Psi}_{ABR}\bra{\Psi})$. The relative entropy of entanglement was defined in Ref. \cite{PhysRevLett.78.2275,PhysRevA.57.1619}, as 
 \begin{align*}
E_R(\rho_{AB})=min_{\sigma}S(\rho_{AB}||\sigma).
 \end{align*} 
 The disturbance of a bipartite channel is defined as 
 \begin{align}\label{eq:9}
 D((\rho_{AB}))&= S(\rho_{AB})- I_c((\rho_{AB}))\nonumber \\
 &= S(\rho_{AB})-S(\mathcal{E}(\rho_{AB})+S(\mathcal{E}\otimes \mathcal{I}(\ket{\Psi}_{ABR}\bra{\Psi}) .
 \end{align} 
 
 Also, the relative entropy of quantum coherence for the bipartite state  can be defined as
  \begin{align}\label{eq:10}
C(\rho_{AB})=S(\rho_{AB}^D)-S(\rho_{AB}).
  \end{align} 
where $\rho_{AB}^D$  is the diagonal part of $\rho_{AB}$ in the basis $\{\ket{i} \otimes\ket{\mu}\} $ $\in $ $ \mathcal{H}_{AB}$.
  Using the above definitions of coherence, entanglement and disturbance for the bipartite state $\rho_{AB}$, we can get a complementarity relation of the following form
\begin{align}
C(\rho_{AB})+E_R(\rho_{AB})+ D(\rho_{AB})\leq2\log(d_{AB}).
\end{align}

The proof of the relation is as follows:
\begin{align*}
&C(\rho_{AB})+E_R(\rho_{AB})+ D(\rho_{AB})\nonumber \\
&=S(\rho_{AB}^D) + min_{\sigma}S(\rho_{AB}||\sigma)-S(\mathcal{E}(\rho_{AB})+S(\mathcal{E}\otimes \mathcal{I}(\ket{\Psi}_{ABR}\bra{\Psi}) \nonumber \\ 
&\leq S(\rho_{AB}^D)+ S(\rho_{AB}||\rho_A\otimes\rho_B) +S(\rho_{AB}) \nonumber \\ 
&=S(\rho_{AB}^D)+S(\rho_A)+S(\rho_B) \\
&\leq 2\log(d_{AB}).
\end{align*}
\noindent
where $d_{AB}$ is the dimension of the state $\rho_{AB}$. The first inequality is obtained using subadditivity of $S(\mathcal{E}\otimes \mathcal{I}(\ket{\Psi}_{ABR}\bra{\Psi})$ and the fact that, $min_{\sigma}S(\rho_{AB}||\sigma) \leq S(\rho_{AB}||\rho_A\otimes\rho_B)$. The final inequality follows from maximum value of the entropy of the states, i.e., $S(\rho_A)\leq\log(d_A)$, $S(\rho_B)\leq\log(d_B)$ and $S(\rho_{AB})\leq\log(d_{AB})$.

\subsection{Complementarity of Coherence, Quantum Discord and Disturbance}
In the last section we proved a complementarity relation for coherence, entanglement and disturbance caused by a CPTP map on a bipartite system. Similarly, one can ask if other quantum correlations like quantum discord satisfies a similar complementarity relation. It was shown in  Ref.\cite{PhysRevLett.116.160407} that for multipartite states, creation of quantum discord with multipartite incoherent operations is bounded by the amount of quantum coherence consumed in its subsystems during the process. This interplay between coherence and quantum discord suggests that coherence, quantum discord and disturbance of a bipartite system could also satisfy a complementarity relation. We will now prove that, they also satisfy a complementarity relation.
Quantum Discord of a bipartite state was defined in Ref.\cite{PhysRevLett.88.017901} as
\begin{align*}
Q_D(\rho_{AB})&= min_{{\Pi}_i^B}[I(\rho_{AB})-J(\rho_{AB})_{{\Pi}_i^B}]\\
&.
\end{align*}

\noindent where $I(\rho_{AB})=S(\rho_A)+S(\rho_B)-S(\rho_{AB})$ is the mutual information between the subsystems $A$ and $B$ and $J(\rho_{AB})_{{\Pi}_i^B}=S(\rho_A)-S(A|{{\Pi}_i^B})$, represents the amount of information gained about the subsystem $A$ by measuring the subsystem $B$. Here, ${\Pi}_i^B$ are the measurement operators corresponding to von neumann measurement on the subsystem $B$. Using the definitions of disturbance and coherence given in Eq.(\ref{eq:9}) and Eq.(\ref{eq:10}) respectively, for a bipartite state, we get a complementarity relation of the following form
\begin{align}
C(\rho_{AB})+Q_D(\rho_{AB})+ D(\rho_{AB})\leq2\log(d_{AB}).
\end{align}

The proof of the relation is as follows:
\begin{align*}
&C(\rho_{AB})+Q_D(\rho_{AB})+ D(\rho_{AB})\nonumber \\
&=S(\rho_{AB}^D) + min_{{\Pi}_i^B}[I(\rho_{AB})-J(\rho_{AB})_{{\Pi}_i^B}]-S(\mathcal{E}(\rho_{AB})+\nonumber \\& \hspace{0.4cm} S(\mathcal{E}\otimes \mathcal{I}(\ket{\Psi}_{ABR}\bra{\Psi}) \nonumber \\ 
&\leq S(\rho_{AB}^D)+ I(\rho_{AB}) +S(\rho_{AB}) \nonumber \\ 
&=S(\rho_{AB}^D)+S(\rho_A)+S(\rho_B) \\
&\leq 2\log(d_{AB}).
\end{align*}

\noindent where the proof is similar to the proof of complementarity relation of entanglement, coherence and disturbance of bipartite state.

\section{Examples}\label{sec:4}
In this section we analyze the coherence disturbance complementarity relation for different quantum channels for a single qubit 
density matrix. The complementarity relations can be neatly presented for few channels. Let us consider a two qubit pure composite state of system and ancilla. 
\begin{align*}
\ket{\Psi}= \sqrt{\lambda}_0\ket{00} + \sqrt{\lambda}_1\ket{11}.
\end{align*}

For this composite state, the density matrix of system  in \{$\ket{+}, \ket{-}$\} basis is given by \\ \\
$\rho= \frac{1}{2}\begin{bmatrix}
1 & \lambda_0-\lambda_1 \\ \lambda_0-\lambda_1 & 1 
\end{bmatrix}$ .\\
\\
\noindent In this basis it has non-zero coherence, which is given by
\begin{align} \label{eq:5}
C_r(\rho)\nonumber 
&= -\mathrm{Tr}[\rho^D \log_2\rho^D]+\mathrm{Tr}[\rho \log_2\rho] \nonumber \\
&= 1 + \frac{1}{2}(1+\lambda_0-\lambda_1)\log_2\frac{1}{2}(1+\lambda_0-\lambda_1) \nonumber \\
&+\frac{1}{2}(1-\lambda_0+\lambda_1)\log_2\frac{1}{2}(1-\lambda_0+\lambda_1).
\end{align}

Disturbance of a state depends on both the state density matrix and the quantum channel. We give Kraus operators and the corresponding expressions of the disturbance and present the complementarity relations, for few channels as examples.

\subsection*{Weak measurement channel}  
	 The theory of weak measurement channel using the measurement operator formalism was done in Ref.\cite{PhysRevLett.95.110409,SINGH2014141,PhysRevA.91.052115}. The approach provides a new tool to handle strong as well weak measurements. The Kraus operators for the weak measurement channel are given by\\   \\
	 $K(x)=\sqrt{\frac{1-x}{2}}\Pi_0 + \sqrt{\frac{1+x}{2}}\Pi_1$, \\
	 $K(-x)  = \sqrt{\frac{1+x}{2}}\Pi_0 + \sqrt{\frac{1-x}{2}}\Pi_1$. 
	 	 
	 \noindent where $\Pi_0$ and $\Pi_1$ are the two projection operators in the computational basis.
	 The weak measurement kraus operators satisfy $K(x)^{\dagger}K(x)+K(-x)^{\dagger}K(-x)=\mathcal{I}$. The parameter $x \in [0,1]$ denotes the measurement strength, where the measurement strength increases as $x$ goes from 0 to 1. The operators satisfy the following properties: $(i)$ For $x=0$, we have no measurement ,i.e., $K(x)$=$K(-x)$$=\frac{\mathcal{I}}{\sqrt{2}}$, resulting in no state change. $(ii)$ For $x=1$, in the strong measurement limit we have the projective measurements, i.e., $K(x)=\Pi_1$ and $K(-x)=\Pi_0$. $(iii)$ $[K{(x)},K{(-x)}]$=0. 	 Under the weak measurement channel the state changes as \\ \\
	 $\rho \to \mathcal{E}(\rho)= \frac{1}{2}\begin{bmatrix}
	 1 & \sqrt{(1-x)}(\lambda_0-\lambda_1) \\ (\sqrt{(1-x)})(\lambda_0-\lambda_1) & 1 
	 \end{bmatrix}.$\\ \\ 

	   	   Disturbance for the weak measurement channel is given by
	   	
\begin{widetext}
		\begin{align} \label{eq:6}
		D(\rho,\mathcal{E}) 
		&= -\mathrm{Tr}\left[ \rho \log_2\rho\right] +\mathrm{Tr}[\mathcal{E}(\rho) \log_2\mathcal{E}(\rho)] 
		- \mathrm{Tr}[\mathcal{E}\otimes I(\ket{\Psi}\bra{\Psi})\log_2\mathcal{E} \otimes I(\ket{\Psi}\bra{\Psi}] \nonumber \\
		&= -\frac{1}{2}(1+\lambda_0-\lambda_1)\log_2\frac{1}{2}(1+\lambda_0-\lambda_1) -\frac{1}{2}(1-\lambda_0+\lambda_1)\log_2\frac{1}{2}(1-\lambda_0+\lambda_1)  \nonumber \\ &+\left(\frac{1}{2}-\frac{\lambda_0-\lambda_1}{2}\sqrt{1-x^2}\right)\log_2\left(\frac{1}{2}-\frac{\lambda_0-\lambda_1}{2}\sqrt{1-x^2}\right)+ \left(\frac{1}{2}+\frac{\lambda_0-\lambda_1}{2}\sqrt{1-x^2}\right)\log_2\left(\frac{1}{2}+\frac{\lambda_0-\lambda_1}{2}\sqrt{1-x^2}\right)\nonumber \\&-\frac{1-\sqrt{1-4\lambda_0\lambda_1x^2}}{2}\log_2\frac{1-\sqrt{1-4\lambda_0\lambda_1x^2}}{2} 
		 -\frac{1+\sqrt{1-4\lambda_0\lambda_1x^2}}{2}\log_2\frac{1+\sqrt{1-4\lambda_0\lambda_1x^2}}{2}.
		\end{align}
\end{widetext}
	 
\begin{figure}
	{\includegraphics[scale=0.36]{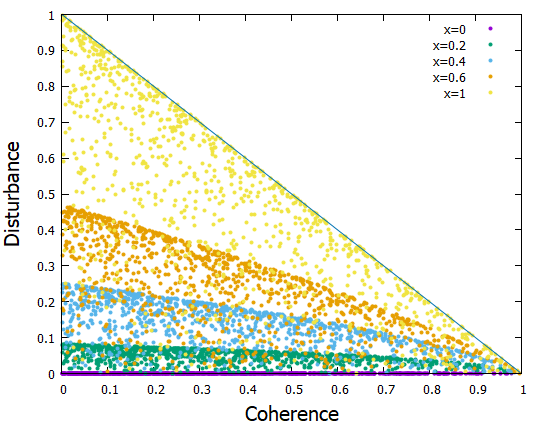}}
	\caption{The complementarity between Coherence $C(\rho)$ and Disturbance $D(\rho,\mathcal{E})$  for the weak measurement channel. The figure shows coherence along the X-axis and disturbance along Y-axis. Random states were generated and coherence and entropy were calculated, using the Matlab package \cite{qetlab}.}
	\label{fig.1}
\end{figure}

It can be checked that $D(\rho,\mathcal{E})$ increases monotonically as $x$ is increased from 0 to 1. By using Eq.(\ref{eq:5}) and Eq.(\ref{eq:6}) we get $C(\rho)+D(\rho,\mathcal{E})\leq1$, plotted in Figure.\ref{fig.1}. This relation is tighter than our original relation Eq.(\ref{eq:3}). The same complementarity relation is also obtained for the bit-flip, phase flip and bit-phase flip channels for a single qubit system.

\subsection*{Depolarizing Channel} 
The Kraus operators for the depolarising channel are given by\\ \\
$	K_1 = \sqrt{1-\frac{3p}{4}} I_2 , K_2 = \sqrt{\frac{p}{4}} \sigma_x ,\\ K_3 = \sqrt{\frac{p}{4}}\sigma_y , K_4 = \sqrt{\frac{p}{4}} \sigma_z$, 

\noindent where $\sigma_x, \sigma_y$ and $\sigma_z$ are the pauli matrices. Under the depolarising channel the state changes as \\ \\
$\rho \to \mathcal{E}(\rho)= \frac{1}{2}\begin{bmatrix}
1 & (1-p)(\lambda_0-\lambda_1) \\ (1-p)(\lambda_0-\lambda_1) & 1 
\end{bmatrix}$ .\\

The disturbance for the depolarising channel is given by\\  

\begin{widetext}
		\begin{align} \label{eq:7}
		D(\rho,\mathcal{E}) 
		&= -\frac{1}{2}(1+\lambda_0-\lambda_1)\log_2\frac{1}{2}(1+\lambda_0-\lambda_1) -\frac{1}{2}(1-\lambda_0+\lambda_1)\log_2\frac{1}{2}(1-\lambda_0+\lambda_1)  +\left(\frac{1}{2}-\frac{\lambda_0-\lambda_1}{2}(1-p)\right)\log_2\left(\frac{1}{2}-\frac{\lambda_0-\lambda_1}{2}(1-p)\right)\nonumber \\
		&+\left(\frac{1}{2}+\frac{\lambda_0-\lambda_1}{2}(1-p)\right)\log_2\left(\frac{1}{2}+\frac{\lambda_0-\lambda_1}{2}(1-p)\right)-\frac{p\lambda_0}{2}\log_2\frac{p\lambda_0}{2} -\frac{p\lambda_1}{2}\log_2\frac{p\lambda_1}{2}\nonumber \\
		&-\left(\frac{\left(1-\frac{p}{2}\right)+\sqrt{(1-\frac{p}{2})^2-4\lambda_0\lambda_1(p-\frac{3p^2}{4})}}{2}\right)\log_2\left(\frac{\left(1-\frac{p}{2}\right)+\sqrt{(1-\frac{p}{2})^2-4\lambda_0\lambda_1(p-\frac{3p^2}{4})}}{2}\right)\nonumber \\&-\left(\frac{\left(1-\frac{p}{2}\right)-\sqrt{(1-\frac{p}{2})^2-4\lambda_0\lambda_1(p-\frac{3p^2}{4})}}{2}\right)\log_2\left(\frac{\left(1-\frac{p}{2}\right)-\sqrt{(1-\frac{p}{2})^2-4\lambda_0\lambda_1(p-\frac{3p^2}{4})}}{2}\right).
		\end{align}
\end{widetext}

\begin{figure}
\includegraphics[scale=0.36]{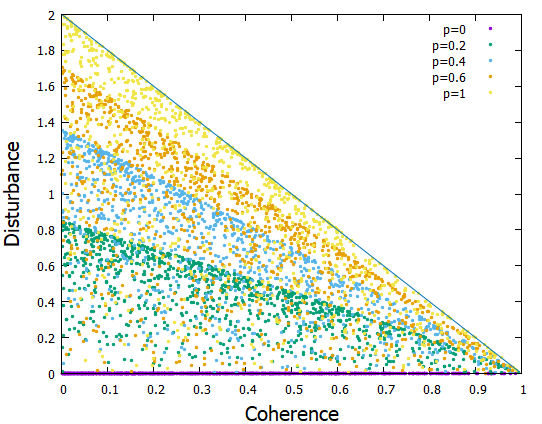}
	\caption{The complementarity between Coherence $C(\rho)$ and Disturbance $D(\rho,\mathcal{E})$  for the depolarising channel. The figure shows coherence along the X-axis and disturbance along Y-axis. Random states were generated and coherence and entropy were calculated, using the Matlab package \cite{qetlab}.}
	\label{fig.2}
\end{figure}

	 Again it is easy to check that $D(\rho,\mathcal{E})$ increases monotonically with $p$. Moreover, using Eq.(\ref{eq:5}) and Eq.(\ref{eq:7}) we get $2C(\rho)+D(\rho,\mathcal{E})\leq2$, which is same as Eq.(\ref{eq:3}), plotted in Figure.\ref{fig.2}.

	\subsection*{Amplitude Damping Channel} 
	
	The Kraus operators for Amplitude damping channel are given by \\ \\
	$	K_1 = \sqrt{q}\ket{0}\bra{1} , K_2 = \ket{0}\bra{0} + \sqrt{1-q}\ket{1}\bra{1}$.\\
\\	Under the amplitude channel the state transforms as \\ 
	$\rho \to \mathcal{E}(\rho)= \frac{1}{2}\begin{bmatrix}
	1 & \sqrt{(1-q)}(\lambda_0-\lambda_1) \\ (\sqrt{(1-q)}+q)(\lambda_0-\lambda_1) & 1-q 
	\end{bmatrix}$. \\ \\
	The disturbance of the amplitude damping channel is given by \\ 
\begin{widetext}
	\begin{align} \label{eq:8}
	D(\rho,\mathcal{E}) 
	&=-\frac{1}{2}(1+\lambda_0-\lambda_1)\log_2\frac{1}{2}(1+\lambda_0-\lambda_1) \nonumber  -\frac{1}{2}(1-\lambda_0+\lambda_1)\log_2\frac{1}{2}(1-\lambda_0+\lambda_1) -(1-q\lambda_1) \log_2(1-q\lambda_1)-q\lambda_1 \log_2 q\lambda_1 \\&+ \frac{1}{2}\left(1-\sqrt{q^2+(\lambda_0-\lambda_1)^2(1-q)} \right)\log_2 \frac{1}{2}\left(1-\sqrt{q^2+(\lambda_0-\lambda_1)^2(1-q)} \right) \nonumber  \\& +\frac{1}{2}\left(1+\sqrt{q^2+(\lambda_0-\lambda_1)^2(1-q)} \right)\log_2 \left(1+\sqrt{q^2+(\lambda_0-\lambda_1)^2(1-q)} \right) 
	\end{align}
\end{widetext}	

\begin{figure} [H]
	\includegraphics[scale=0.36]{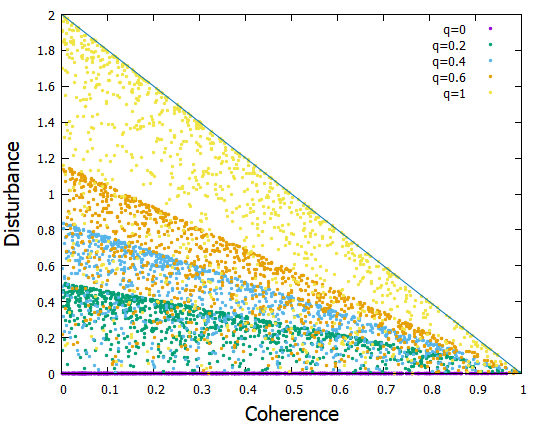}
	\caption{The complementarity between Coherence $C(\rho)$ and Disturbance $D(\rho,\mathcal{E})$  for the amplitude damping channel. The figure shows coherence along the X-axis and disturbance along Y-axis. Random states were generated and coherence and entropy were calculated, using the Matlab package \cite{qetlab}.}
	\label{fig.3}
\end{figure}

	For the amplitude damping channel also $D(\rho,\mathcal{E})$ and $C(\rho)$ follow  the original relation Eq.(\ref{eq:3}). The complementarity relations derived above can be verified with the given plotted in Figure.\ref{fig.3}.

From the figures we note that the complementarity relation for the coherence and disturbance given in Eq.(\ref{eq:3}) is satisfied for all the above channels for single qubit systems. The amount of disturbance reduces as the measurement strength is decreased which is expected in the case of all the channels. It can be also seen that, the complementarity between coherence and disturbance is channel dependent. The complementarity relation obeyed for a single qubit state in case of  weak measurement channel is tighter than Eq.(\ref{eq:3}). While the amplitude damping and depolarising channels follow the original relation  for a single qubit state. In addition to qubit states we found numerically, that for a single qutrit state, the depolarizing channel satisfies a complementarity relation while the amplitude damping channel tightly satisfies the original complementarity relation. 

\section{Conclusion}\label{sec:5}
To summarize, we show that there exists a complementarity relation between the coherence of a state and disturbance caused by a CPTP map or a measurement channel on a quantum system. For measurement channel we find a tighter complementarity relation. Moreover, we obtain a complementarity relation for the quantum coherence, relative entropy of entanglement and disturbance for a bipartite system. Similar relation is also obtained for the quantum coherence, quantum discord and disturbance for a bipartite state. The complementarity relation for the coherence and disturbance has been illustrated for weak measurement channel and other quantum channels. Our results capture the intuition 
that coherence, entanglement and quantum discord for a quantum system, should respect a trade-off relation with disturbance. We hope that these results will find interesting applications 
where we send single or composite systems under noisy channels that tend to loose quantum coherence and entanglement. If we wish to maintain coherence or entanglement or both, then 
we need to send the quantum states through a channel that does not disturb the system to a greater extent. In future it will be interesting to see if other measures of coherence and entanglement respects the complementarity relation with disturbance.

\section*{ACKNOWLEDGEMENT}
GS would like to acknowledge the research fellowship of Department of Atomic Energy, Government of India.

\bibliographystyle{apsrev4-1}
\bibliography{cohdisref}

\end{document}